\documentstyle[12pt]{article}
\begin{document}
\thispagestyle{empty}
\newcommand{\p}[1]{(\ref{#1})}
\newcommand{\be}{\begin{equation}}
\newcommand{\ee}{\end{equation}}
\newcommand{\sect}[1]{\setcounter{equation}{0}\section{#1}}
\newcommand{\B}{{\cal B}}
\newcommand{\F}{{\cal F}}
\newcommand{\vs}[1]{\rule[- #1 mm]{0mm}{#1 mm}}
\newcommand{\hs}[1]{\hspace{#1mm}}
\newcommand{\mb}[1]{\hs{5}\mbox{#1}\hs{5}}
\newcommand{\Db}{{\overline D}}
\newcommand{\bea}{\begin{eqnarray}}
\newcommand{\eea}{\end{eqnarray}}
\newcommand{\wt}[1]{\widetilde{#1}}
\newcommand{\und}[1]{\underline{#1}}
\newcommand{\ov}[1]{\overline{#1}}
\newcommand{\sm}[2]{\frac{\mbox{\footnotesize #1}\vs{-2}}
           {\vs{-2}\mbox{\footnotesize #2}}}
\newcommand{\prt}{\partial}
\newcommand{\eps}{\epsilon}
\newcommand{\R}{\mbox{\rule{0.2mm}{2.8mm}\hspace{-1.5mm} R}}
\newcommand{\Z}{Z\hspace{-2mm}Z}
\newcommand{\cd}{{\cal D}}
\newcommand{\cg}{{\cal G}}
\newcommand{\ck}{{\cal K}}
\newcommand{\cw}{{\cal W}}
\newcommand{\vj}{\vec{J}}
\newcommand{\vl}{\vec{\lambda}}
\newcommand{\vz}{\vec{\sigma}}
\newcommand{\vt}{\vec{\tau}}
\newcommand{\vw}{\vec{W}}
\newcommand{\poiss}{\stackrel{\otimes}{,}}
\def\l#1#2{\raisebox{.2ex}{$\displaystyle
  \mathop{#1}^{{\scriptstyle #2}\rightarrow}$}}
\def\r#1#2{\raisebox{.2ex}{$\displaystyle
 \mathop{#1}^{\leftarrow {\scriptstyle #2}}$}}
\newcommand{\NP}[1]{Nucl.\ Phys.\ {\bf #1}}
\newcommand{\PL}[1]{Phys.\ Lett.\ {\bf #1}}
\newcommand{\NC}[1]{Nuovo Cimento {\bf #1}}
\newcommand{\CMP}[1]{Comm.\ Math.\ Phys.\ {\bf #1}}
\newcommand{\PR}[1]{Phys.\ Rev.\ {\bf #1}}
\newcommand{\PRL}[1]{Phys.\ Rev.\ Lett.\ {\bf #1}}
\newcommand{\MPL}[1]{Mod.\ Phys.\ Lett.\ {\bf #1}}
\newcommand{\BLMS}[1]{Bull.\ London Math.\ Soc.\ {\bf #1}}
\newcommand{\IJMP}[1]{Int.\ Jour.\ of\ Mod.\ Phys.\ {\bf #1}}
\newcommand{\JMP}[1]{Jour.\ of\ Math.\ Phys.\ {\bf #1}}
\newcommand{\LMP}[1]{Lett.\ in\ Math.\ Phys.\ {\bf #1}}
\renewcommand{\thefootnote}{\fnsymbol{footnote}}
\newpage
\setcounter{page}{0}
\pagestyle{empty} 
\begin{flushright}
{November 1997}\\
{SISSA 142/97/EP}\\
{solv-int/9711009}
\end{flushright}
\vs{8}
\begin{center}
{\LARGE {\bf The $N=2$ supersymmetric}}\\[0.6cm]
{\LARGE {\bf matrix GNLS hierarchies}}\\[1cm]

\vs{8}

{\large L. Bonora$^{a,1}$, S. Krivonos$^{b,2}$ and A. Sorin$^{b,3}$}
{}~\\
\quad \\
{\em ~$~^{(a)}$ International School for Advanced Studies (SISSA/ISAS),}\\
{\em Via Beirut 2, 34014 Trieste, Italy}\\
{\em INFN, Sezione di Trieste}\\
{\em {~$~^{(b)}$ Bogoliubov Laboratory of Theoretical Physics, JINR,}}\\
{\em 141980 Dubna, Moscow Region, Russia}~\quad\\

\end{center}
\vs{8}

\centerline{ {\bf Abstract}}
\vs{4}

We construct the matrix generalization of the $N=2$ supersymmetric 
GNLS hierarchies. 
This is done by exhibiting the corresponding matrix super Lax operators
in terms of $N=2$ superfields in two different superfield bases.
We present the second Hamiltonian structure and discrete symmetries.
We then extend our discussion by conjecturing the Lax operators 
of different reductions of the $N=2$ supersymmetric matrix KP hierarchy and 
discuss the simplest examples.

\vfill
{\em E-Mail:\\
1) bonora@sissa.it\\
2) krivonos@thsun1.jinr.dubna.su\\
3) sorin@thsun1.jinr.dubna.su }
\newpage

\pagestyle{plain}
\renewcommand{\thefootnote}{\arabic{footnote}}
\setcounter{footnote}{0}

\noindent{\bf 1. Introduction.}
The $N=2$ supersymmetric $(n,m)$--Generalized Nonlinear Schr\"{o}dinger
(GNLS) hierarchies were introduced in \cite{bks}. We recall that they form
a very large family of $N=2$ supersymmetric hierarchies, one out three
known families of $N=2$ supersymmetric hierarchies with $N=2$ $W_s$ second
 Hamiltonian structure. The $N=2$ $(n,m)$--GNLS hierarchies were subsequently 
studied in a number of papers, [2--9]. In
particular 
their discrete symmetries were derived in \cite{s1}, and the Hamiltonian 
structures and recursion operator were constructed in \cite{bs} in two 
different superfield bases with local evolution equations. The $N=2$
$(n,m)$--GNLS hierarchy involves $n+m$ pairs of chiral and antichiral 
$N=2$ superfields,
$n$ pairs of them being bosonic and $m$ pairs fermionic.
In order to define the matrix generalization, we combine them into a single 
row and a single column of $(n+m)$ length. The aim of the present letter 
is to generalize the $N=2$
$(n,m)$--GNLS hierarchy to the case when the row and column are
replaced by a rectangular matrix of an arbitrary $(k\times(n+m))$--size and
its transposed matrix, respectively. It appears that such integrable
generalization actually exists, and many results of Refs. \cite{s1,bs}
concerning the $N=2$ super $(n,m)$--GNLS hierarchy can be straightforwardly
extended to this case. This permits to present here the main facts
concerning the new hierarchy, called $N=2$ supersymmetric $(k|n,m)$
matrix GNLS (MGNLS) hierarchy, in a telegraphic style and refer the reader to
Refs. \cite{bks,s1,bs} for more details. In section 2 we introduce the
new matrix hierarchies by means of their Lax operators. In section 3 we
present their Hamiltonian structures. In section 4 we write the same 
hierarchies in a new basis, the so--called KdV basis, while in section 5
we discuss the discrete symmetries of these hierarchies. Section 6 is devoted
to a generalization: we start from the $N=2$ matrix KP hierarchy 
(which contains an infinite number of fields) and conjecture the existence of
an infinite family of reductions (with a finite number of fields), which 
includes in particular the $N=2$ matrix GNLS hierarchies as well as almost
all known $N=2$ scalar KP reductions.

{}~

\noindent{\bf 2. The $N=2$ super $(k|n,m)$-MGNLS hierarchy.}
The Lax operator of the $N=2$ supersymmetric MGNLS
hierarchies has the following form
\begin{eqnarray}
L= I\partial - \frac{1}{2}(F {\overline F} +
F {\overline D}\partial^{-1} \left[ D {\overline F}\right]), \quad
[D,L]=0.
\label{suplax}
\end{eqnarray}
Here, $F\equiv F_{Aa}(Z)$ and ${\overline F}\equiv {\overline F}_{aA}(Z)$
($A,B=1,\ldots, k$; $a,b=1,\ldots , n+m$) are chiral and antichiral
rectangular matrix-valued $N=2$ superfields,
\begin{eqnarray}
D F=0, \quad {\overline D}~{\overline F} = 0,
\label{chiral}
\end{eqnarray}
respectively. In (\ref{suplax}) the matrix product is understood, for example 
$(F\overline F)_{AB} \equiv \sum_a F_{Aa}\overline F_{aB}$. Moreover  
the square brackets mean that the relevant operators act only on the superfields
inside the brackets, and $I$ is the unity matrix, $I\equiv {\delta}_{A,B}$.
The matrix entries are bosonic superfields for $a=1,\ldots ,n$ and
fermionic superfields for $a=n+1,\ldots , n+m$, i.e.,
$F_{Aa}{\overline F}_{bB}=(-1)^{d_{a}{\overline d}_{b}}{\overline
F}_{bB}F_{Aa}$, where $d_{a}$ and ${\overline d}_{b}$ are the Grassmann
parities of the matrix elements $F_{Aa}$ and ${\overline F}_{bB}$,
respectively, $d_{a}=1$ $(d_{a}=0)$ for fermionic (bosonic) entries;
$Z=(z,\theta,\overline\theta)$ is a coordinate of the $N=2$ superspace,
$dZ \equiv dz d \theta d \overline\theta$ and $D,{\overline D}$ are the
$N=2$ supersymmetric fermionic covariant derivatives
\begin{eqnarray}
D=\frac{\partial}{\partial\theta}
 -\frac{1}{2}\overline\theta\frac{\partial}{\partial z}, \quad
{\overline D}=\frac{\partial}{\partial\overline\theta}
 -\frac{1}{2}\theta\frac{\partial}{\partial z}, \quad
D^{2}={\overline D}^{2}=0, \quad
\left\{ D,{\overline D} \right\}= -\frac{\partial}{\partial z}
\equiv -{\partial}.
\label{DD}
\end{eqnarray}
Let us stress that the chosen grading guarantees that the Lax operator $L$ is
Grassman even, and it is consistent with the important property that
$L$ commutes with fermionic derivative $D$ (see, eqs. \p{suplax}).

For $p=0,1,2,.. $, the Lax operator $L$ provides the consistent flows
\begin{eqnarray}
{\textstyle{\partial\over\partial t_p}}L =[ A_p , L],\quad
A_p=(L^p)_{\geq 1}.
\label{laxfl1}
\end{eqnarray}
For the particular case $k=1$, the Lax operator \p{suplax} coincides with
the scalar Lax operator of the $N=2$ super $(n,m)$--GNLS hierarchy while for
$k \geq 2$ it defines the new $N=2$ super $(k|n,m)$--MGNLS
hierarchy. Like in the scalar case, the infinite number of
Hamiltonians can be obtained as follows:
\begin{eqnarray}
{H}_p =\int d Z {\cal H}_p, \quad  {\cal H}_p \equiv tr(L^p)_{0},
\label{res}
\end{eqnarray}
where the subscripts $\geq 1$ and $0$ denote the sum of the purely
derivative terms and the constant part of the operator, respectively, and
$tr$ means the usual matrix trace. The evolution
equations for the $F$ and $\overline F$ derived from \p{laxfl1} 
admit the involution
\begin{eqnarray}
F^{*}= i{\overline F}^{T}{\cal I}, \quad
{\overline F^{*}}=i{\cal I} F^{T}, \quad
{\theta}^{*}={\overline {\theta}}, \quad
{\overline {\theta}^{*}}={\theta}, \quad
t^{*}_p= (-1)^{p+1} t_p, \quad z^{*}= z,
\quad i^{*}=-i,
\label{conj}
\end{eqnarray}
where $i$ is the imaginary unit, the symbol $T$ means the operation of
matrix transposition, and the matrix ${\cal I}$ is
\begin{eqnarray}
{\cal I}\equiv (-i)^{d_a} {\delta}_{ab},
\label{matrI}
\end{eqnarray}
with the properties
\begin{eqnarray}
{\cal I}{\cal I}^{*}=I, \quad {\cal I}^{3}={\cal I}^{*},\quad
{\cal I}^2= (-1)^{d_a}{\delta}_{ab}.
\label{matrII}
\end{eqnarray}
The formulae \p{conj}--\p{matrII} provide the following relation
\begin{eqnarray}
(F{\overline F})^{*}= -(F{\overline F})^{T},
\label{conjrel}
\end{eqnarray}
which is exploited in what follows. The origin of the involution \p{conj}
is explained in the next section.

The flows \p{laxfl1} are local, and can be represented in the following
form:
\begin{eqnarray}
{\textstyle{\partial\over\partial t_p}}F =
((L^p)_{\geq 1} F)_0, \quad {\textstyle{\partial\over\partial
t_p}}\overline F =(-1)^{p+1} ({\overline F}(\r {{L^\dagger}^p}{})_{\geq 1})_0,
\label{n1}
\end{eqnarray}
where the arrows mean that the corresponding operators act on the left
and $\dagger$ denotes hermiten conjugation, i.e. involution plus transposition,
\begin{eqnarray}
\r {L^\dagger}{} \equiv (\r {L^*}{})^{T}= I\r {\partial}{} +
\frac{1}{2}(F{\overline F} +
\left[ {\overline D}F\right] \r D{}\r {{\partial}^{-1}}{}{\overline F}) ,
\quad [~\overline D, \r {L^{\dagger}}{}]=0.
\label{suplaxconj}
\end{eqnarray}
This Lax operator also provides consistent flows. 

The first three flows from \p{n1} and the first three nontrivial
Hamiltonian densities from \p{res} are:
\begin{eqnarray}
&&{\textstyle{\partial\over\partial t_0}} F = F,\quad
{\textstyle{\partial\over \partial t_0}} {\overline F} =
-{\overline F}; \quad
{\textstyle{\partial\over\partial t_1}} F =F~', \quad
{\textstyle{\partial\over \partial t_1}} {\overline F} =
{\overline F}~';\nonumber\\
&&{\textstyle{\partial\over\partial t_2}} F =
F~'' +  {D}(F {\overline F}~{\overline D} F), \quad
{\textstyle{\partial\over \partial t_2}} {\overline F} =
-{\overline F}~'' + {\overline D}([{D}{\overline F}]F {\overline F}),
\label{gnls}
\end{eqnarray}
\begin{eqnarray}
{\cal H}_1 &=& -\frac{1}{2} tr (F {\overline F}), \quad
{\cal H}_2 = \frac{1}{2}tr ( F {\overline F}~' +
\frac{1}{4} (F{\overline F})^2), \nonumber\\
{\cal H}_3 &=& -\frac{1}{2}tr ( F {\overline F}~'' -
\frac{1}{2} \left[{\overline D} F {\overline F}\right]
\left[D F {\overline F}\right] +
F {\overline F}~' F {\overline F} + \frac{1}{12} (F {\overline F})^3),
\label{i1}
\end{eqnarray}
respectively, where $'$ means the derivative with respect to $z$.

Now we consider the bosonic limit of the second flow equations \p{gnls}
both for the case of pure fermionic ($n=0$) and pure bosonic ($m=0$) matrices
$F, {\overline F}$, and establish their relations with the $gl(k+m)/(gl(k)
\times gl(m))$ bosonic matrix NLS equations introduced in \cite{fk}.

To derive the bosonic limit, let us define the matrix components of the
fermionic superfield matrices
\begin{eqnarray}
f = {\overline D} F|, \quad {\overline f} = D {\overline F}|, \quad
{\psi} =  F|, \quad {\overline {\psi}} ={\overline F}|,
\end{eqnarray}
and the matrix components of the bosonic superfield matrices
\begin{eqnarray}
{\xi} = {\overline D} F|, \quad {\overline {\xi}}=D {\overline F}|, \quad
b = F|, \quad {\overline b} ={\overline F}|,
\end{eqnarray}
where $|$ means the $({\theta}, {\bar\theta})\rightarrow 0$
limit. So, $\psi,{\overline \psi},\xi, {\overline \xi}$ are fermionic
matrix components while $f,{\overline f},b, {\overline b}$ are
bosonic ones. To get the bosonic limit we have to put all the fermionic
matrices ${\psi}, {\overline {\psi}}, {\xi}, {\overline {\xi}}$ to zero.
This leaves us with the following set of matrix equations
\begin{eqnarray}
{\textstyle{\partial\over\partial t_2}} f=f~''- f{\overline f}f, \quad
{\textstyle{\partial\over \partial t_2}} {\overline f} =
-{\overline f}~''+{\overline f}f{\overline f},
\label{bosgnls1}
\end{eqnarray}
\begin{eqnarray}
{\textstyle{\partial\over\partial t_2}} b=b~''- b{\overline b} b~',\quad
{\textstyle{\partial\over \partial t_2}} {\overline b}=
-{\overline b}~''-{\overline b}~'b{\overline b}
\label{newgnls1}
\end{eqnarray}
for the bosonic matrix components. The set of equations \p{bosgnls1} form the
bosonic matrix NLS equations which can be produced
via $gl(k+m)/(gl(k)\times gl(m))$-coset construction \cite{fk}. They
can be viewed as the second flow of the bosonic matrix NLS hierarchies with the
Lax operators ${\cal L}_1$
\begin{eqnarray}
{\cal L}_1= I\partial - \frac{1}{2}f \partial^{-1} {\overline f}, \quad
{\textstyle{\partial\over\partial t_p}}{\cal L}_1 =
[ ({\cal L}_1^p)_{\geq 0} , L].
\label{boslax2}
\end{eqnarray}
The latter can be easily derived from the Lax operator \p{suplax} in the
bosonic limit\footnote{For more details concerning bosonic matrix NLS 
equations, see \cite{hw,ly} and references therein.}. In the same way one
can derive the Lax operator for the equations \p{newgnls1},
\begin{eqnarray}
{\cal L}_2= 
(I -\frac{1}{2} b \partial^{-1} {\overline b}) {\partial} ,
\label{boslax1}
\end{eqnarray}
which generates the hierarchy which we call modified matrix NLS hierarchy.

Thus we can conclude that the $N=2$ supersymmetric $(k|n,m)$--MGNLS
hierarchies with Lax operators \p{suplax} are the $N=2$ supersymmetric
extensions of the bosonic matrix NLS and modified matrix NLS hierarchies.
In a companion paper \cite{bks2} we show that the $N=2$ 
$(k|n,m)$--MGNLS hierarchies can be obtained via a suitable coset construction
applied to the $sl(s|s-1)$ $N=2$ affine superalgebras.

{}~

\noindent{\bf 3. Hamiltonian structure of the $N=2$ super
$(k|n,m)$-MGNLS hierarchy.}
The system of evolution equations \p{gnls} is Hamiltonian and can be
represented as
\begin{eqnarray}
{\textstyle{\partial\over\partial t_p}}
\left(\begin{array}{cc} F_{Aa}\\{\overline F}_{aA} \end{array}\right) =
(J_2)_{Aa,Bb}\left(\begin{array}{cc} {\delta}/{\delta F_{Bb}} \\
{\delta}/{\delta {\overline F}_{bB}} \end{array}\right) H_{p}, 
\label{hameq}
\end{eqnarray}
where summation over repeated indices is understood, and $J_2$ is the
second Hamiltonian structures, which has the following form:
\begin{eqnarray}
&& (J_2)_{Aa,Bb}= \left(\begin{array}{cc} (J_{11})_{Aa,Bb}, &
(J_{12})_{Aa,Bb} \\
(J_{21})_{Aa,Bb} &  (J_{22})_{Aa,Bb}
\end{array}\right),  \nonumber\\
&& (J_{11})_{Aa,Bb}=(-1)^{d_ad_b} F_{Ab} D{\overline D} \partial^{-1}
F_{Ba}- F_{Ba} D{\overline D} \partial^{-1} F_{Ab}, \nonumber\\ &&
(J_{12})_{Aa,Bb}=(-1)^{d_b}(2D \overline D {\delta}_{AB} -
F_{Ac}D {\overline D}\partial^{-1}{\overline F}_{cB})
{\delta}_{ab}+ F_{Ca} D{\overline D} \partial^{-1}
{\overline F}_{bC}{\delta}_{AB}, \nonumber\\ && (J_{21})_{Aa,Bb}=
(2{\overline D} D {\delta}_{AB}+(-1)^{d_c}{\overline F}_{cA}~
{\overline D}D\partial^{-1} F_{Bc}){\delta}_{ab} -{\overline F}_{aC}~
{\overline D}D\partial^{-1}F_{Cb}{\delta}_{AB}, \nonumber\\
&& (J_{22})_{Aa,Bb}= {\overline F}_{aB}~{\overline D}D{\partial^{-1}}
{\overline F}_{bA} - (-1)^{d_ad_b}{\overline F}_{bA}~
{\overline D}D {\partial^{-1}}{\overline F}_{aB}.
\label{hamstr2}
\end{eqnarray}
This formula should be expressible in a more compact way by using the 
$r$--matrix language, but we postpone this development to another occasion.

In terms of $J_2$, the $N=2$ supersymmetric
Poisson brackets algebra of the matrices $F$ and ${\overline F}$
are given by the formula:
\begin{eqnarray}
\{\left(\begin{array}{cc}
F_{Aa}(Z_1)\\{\overline F_{aA}(Z_1)}\end{array}\right)
\stackrel{\otimes}{,}
\left(\begin{array}{cc} F_{Bb}(Z_2),{\overline F_{bB}(Z_2)}
\end{array}\right)\}_2=(J_2)_{Aa,Bb}(Z_1){\delta}^{N=2}(Z_1-Z_2),
\label{palg}
\end{eqnarray}
where ${\delta}^{N=2}(Z) \equiv \theta {\overline \theta} {\delta}(z)$ is
the delta function in the $N=2$ superspace and the notation `$\otimes$' stands
for the tensor product. $J_2$ satisfies the Jacobi identities and the symmetry
properties related to the statistics of the matrix entries. In addition 
$J_2$ also satisfies the chiral consistency conditions
\begin{eqnarray} J_2
\Pi={\overline \Pi} J_2=0, \quad J_2 {\overline \Pi}= \Pi J_2=J_2,
\label{cons}
\end{eqnarray}
where we introduced the matrices $\Pi$ and ${\overline {\Pi}}$
\begin{eqnarray}
\Pi & \equiv & - \left(\begin{array}{cc} D\overline D \partial^{-1}
{\delta}_{ab}{\delta}_{AB}, & 0 \\
0, & \overline D D \partial^{-1}{\delta}_{ab}{\delta}_{AB}
\end{array}\right), \quad
{\overline {\Pi}} \equiv - \left(\begin{array}{cc} {\overline D}
D \partial^{-1}{\delta}_{ab}{\delta}_{AB}, & 0 \\
0, & D{\overline D}  \partial^{-1}{\delta}_{ab}{\delta}_{AB}
\end{array}\right), \nonumber\\
&& \Pi \Pi =\Pi, \quad {\overline \Pi}~ {\overline \Pi}=\overline \Pi,
\quad \Pi \overline \Pi=\overline \Pi \Pi=0, \quad \Pi + \overline \Pi =I
\label{pi}
\end{eqnarray}
which project on
the chiral/antichiral and antichiral/chiral subspaces, respectively.
Eqs.\p{cons} show that the second Hamiltonian structure $J_2$ is
represented by a  degenerate matrix. One should stress that this is not a
pathology of the Hamiltonian structure but a peculiarity of the $N=2$
superfield description, which can be easily dealt with, see \cite{bs}.

Let us turn now to the involution properties announced in the previous section.
Under the action of the involution \p{conj} the
Poisson brackets algebra \p{hamstr2}--\p{palg} change the overall sign  while
the Hamiltonians $H_p$ \p{i1} trasform as
\begin{eqnarray}
H^{*}_p= (-1)^{p}H_p,
\label{conjrel1}
\end{eqnarray}
as one can check. This shows that the Hamiltonian system \p{hameq} of 
the $N=2$ $(k|n,m)$-MGNLS flows is invariant under the involution \p{conj}, as  
announced above.

Let us note that besides the Hamiltonians $H_p$ \p{i1}, which are in
involution with respect to the Poisson structure \p{hamstr2}--\p{palg},
there are integrals of the flows \p{laxfl1}, \p{hameq}, which form a
non-abelian algebra. Some of them are matrix-valued local integrals, 
for example,  
\begin{eqnarray}
H_{1,ab} = \int d Z ({\overline F} F)_{ab}.
\label{nres0}
\end{eqnarray}
For the flows \p{gnls}, this can be checked by direct calculations.
Using the algebra \p{palg} one can calculate the Poisson brackets between
them and in this way generate new integrals.
By repeatedly applying the same procedure, one can produce new series of both
local and nonlocal matrix-valued integrals, \cite{bs}.

{}~

\noindent{\bf 4. The $N=2$ super $(k|n,m)$-MGNLS hierarchy
in the KdV-basis.}
One can rewrite the Lax operator \p{suplax} in a slightly
different, but equivalent, form,
\begin{eqnarray}
L= I\partial - \frac{1}{2}(B {\overline B} +
B{\overline D}\partial^{-1} \left[ D {\overline B}\right]+S{\overline S}+
S{\overline D}\partial^{-1} \left[ D {\overline S}\right]),
\label{suplax1}
\end{eqnarray}
where $B\equiv B_{AC}$ and ${\overline B}\equiv {\overline B}_{AC}$ are purely
bosonic or purely fermionic square-matrix extracted from the initial matrices 
$F$ and ${\overline F}$, respectively, and $S\equiv F-B$ and  $\overline S
\equiv \overline F - \overline B$. So $B$ and $\overline B$ are $k\times k$
matrices, while $S$ ($\overline S$) is a $k\times (n+m-k)$ 
($(n+m-k)\times k$) matrix. This can always be done
for the bosonic (fermionic) matrices $B$ and ${\overline B}$ provided
$n \geq k$ ($m \geq k$). Actually there is an ambiguity, for instance, 
in the choice of the $k$ columns among $n$ columns (when $k<n$) that form 
the matrix $B$. However such ambiguity is irrelevant as it corresponds to
an internal symmetry of the Lax operator. In the following let us consider 
the two cases $n \geq k$ and $m \geq k$ separately.

We start with the case when the matrices $B$ and ${\overline B}$ are
bosonic ones. We call them $\B$ and $\overline \B$, respectively. 
We apply the gauge transformation
\begin{eqnarray}
L^{KdV} = \B^{-1} L \B, \quad A^{KdV}_p =
\B^{-1} A_p \B - \B^{-1} {\textstyle{\partial\over\partial t_p}} \B, \quad
{\textstyle{\partial\over\partial t_p}}L^{KdV} = [A^{KdV}_p , L^{KdV}]
\label{n2}
\end{eqnarray}
and substitute the $t_p$-derivative of $\B$ obtained from \p{n1} into
\p{n2}. Then, introducing the new basis
$\{J\equiv J_{AC},\Phi \equiv {\Phi}_{Aa},
{\overline \Phi}\equiv {{\overline \Phi}}_{aA};
A=1,\ldots, k; a=n-k, \ldots , n, \ldots , n+m\}$
\begin{eqnarray}
J^{T}=\frac{1}{2}\B^{-1}(\frac{1}{2} \B {\overline \B}+
\frac{1}{2} S {\overline S} - \partial)\B, \quad
{\Phi}^{T} = \frac{1}{\sqrt{2}} D ({\overline S}\B), \quad
{\overline \Phi}^{T}=\frac{1}{\sqrt{2}} \overline D ({\B}^{-1}S),
\label{n3}
\end{eqnarray}
and making obvious algebraic manipulations in the result, we obtain the
following explicit expressions for the operators $L^{KdV}$ and $A^{KdV}_p$:
\begin{eqnarray}
&&L^{KdV} =
I\partial - 2J^{T}-2 {\overline D}\partial^{-1}
\left[ D (J -\frac{1}{2}({\Phi}
\partial^{-1}{\overline {\Phi}})^{T})\right]
+ \left[ D \partial^{-1} {\overline {\Phi}}^{T}\right]
{\overline D}\partial^{-1} {\Phi}^{T},\nonumber\\
&&A^{KdV}_p=(L^{KdV}_p)_{\geq 1}.\label{n4}
\end{eqnarray}
The flows \p{gnls} and Hamiltonian densities \p{i1} now become
\begin{eqnarray}
&&{\textstyle{\partial\over\partial t_0}}J=
{\textstyle{\partial\over\partial t_0}}{\overline \Phi}=
{\textstyle{\partial\over\partial t_0}}\Phi=0; \quad
{\textstyle{\partial\over\partial t_1}}J=J~',\quad
{\textstyle{\partial\over\partial t_1}}{\overline \Phi}=
{\overline \Phi}~', \quad
{\textstyle{\partial\over\partial t_1}}\Phi_i=\Phi~'; \nonumber\\
&&{\textstyle{\partial\over\partial t_2}}{J} =
(-[D,{\overline D}~] J - 2 J^2+{\Phi}{\overline \Phi})~'-
2[J,[D,{\overline D}~] J-{\Phi}{\overline \Phi}], \nonumber\\
&&{\textstyle{\partial\over\partial t_2}}{\Phi} =
-{\Phi}~''+ 4D\overline D(J\Phi), \quad
{\textstyle{\partial\over\partial t_2}}{\overline \Phi} =
{\overline \Phi}~''+ 4\overline D D ({\overline \Phi} J),
\label{newgnls}
\end{eqnarray}
\begin{eqnarray}
{\cal H}_1 = -2 tr (J), \quad
{\cal H}_2 =tr (  2 J^2 -\Phi {\overline \Phi}), \quad
{\cal H}_3 = tr ( {\Phi}~'~{\overline \Phi} +
4 J \Phi {\overline \Phi} - 4 {\overline D}J DJ - \frac{8}{3} J^3),
\label{i2}
\end{eqnarray}
respectively\footnote{Let us recall that Hamiltonian densities
are defined up to terms which are fermionic or bosonic total
derivatives of arbitrary nonsingular, local functions of the
superfield matrices.}, where the brackets $[,]$ represent the commutator. 
In addition to the first involution \p{conj} hidden in
this basis, they admit an extra, second involution
\begin{eqnarray}
{\Phi}^{*}={\overline \Phi}^{T}{\cal I},\quad
{\overline \Phi}^{*}={\cal I}{\Phi}^{T}, \quad J^{*} = -J^{T}, \quad
{\theta}^{*}={\overline {\theta}}, \quad
{\overline {\theta}^{*}}={\theta}, \quad
t^{*}_p= (-1)^{p+1} t_p, \quad z^{*}= z,
\label{nnconj}
\end{eqnarray}
which is manifest in this basis, but is hidden in the former one.  We
call the basis \p{n3} a KdV--basis, because in the scalar case it 
coincides with the KdV--basis introduced in \cite{bs}. In the KdV-basis, 
the $N=2$ $(k|n,m)$-MGNLS hierarchy of integrable equations, 
together with its
Hamiltonians, can be calculated using formulas \p{laxfl1}, \p{res}, where
the Lax operator $L$ \p{suplax} is replaced by the gauge related Lax
operator $L^{KdV}$ \p{n4}.

The second Hamiltonian structures $J^{KdV}_2$ in the KdV-basis are
related to $J_2$ \p{hamstr2} by the general rule\footnote{Let us recall
the rules for the adjoint conjugation operation ${\cal T}$: $D^{{\cal
T}}=-D$, ${\overline D}^{{\cal T}}= -{\overline D}$, $(QP)^{{\cal T}}=
(-1)^{d_Qd_P}P^{{\cal T}}Q^{{\cal T}}$, where $Q$ and $P$ are arbitrary
operators. For matrices, this operation means the matrix
transposition. All other rules can be derived using these.}
\begin{eqnarray}
J^{KdV}_2= {\cal G} J_2 {\cal G}^{\cal T},
\label{tran}
\end{eqnarray}
where ${\cal G}$ is the matrix of Fr\'echet derivatives corresponding to the
transformation $\{\B,{\overline \B}, S,{\overline S}\}$$\Rightarrow$
$\{J,{\overline \Phi},\Phi\}$ 
 \p{n3} to the KdV-basis.
The calculation of $J^{KdV}_2$ via formula \p{tran} is a simple
exercise and we do not reproduce it here.

Now, let us consider the second case, i.e., when the matrices $B$ and
${\overline B}$ are fermionic ones. We relabel them $\F$ and $\overline \F$,
respectively. Then introducing the new basis 
$\{J\equiv J_{AC},\Phi \equiv {\Phi}_{Aa},
{\overline \Phi}\equiv {{\overline \Phi}}_{aA};
A=1,\ldots, k; a=1, \ldots , n, \ldots , n+m-k\}$
\begin{eqnarray}
J=-\frac{1}{2}[D{\overline \F}] (\frac{1}{2} \F {\overline \F} + \frac{1}{2}
S {\overline S} - \partial)[D{\overline \F}]^{-1}, \quad
\Phi=\frac{1}{\sqrt{2}}D ({\overline \F}S), \quad {\overline \Phi}=
\frac{1}{\sqrt{2}}{\overline D}(D{\overline S}[D {\overline \F}]^{-1}),
\label{nn3}
\end{eqnarray}
and applying the gauge transformation
\begin{eqnarray}
L^{KdV} = [D {\overline \F}] L [D {\overline \F}]^{-1} \equiv
\partial + 2J-2\left[ D (J - \frac{1}{2}{\Phi}
\partial^{-1} {\overline \Phi})\right]
{\overline D}\partial^{-1} + \Phi {\overline D}\partial^{-1}
\left[ D \partial^{-1} {\overline \Phi} \right]
\label{nn2}
\end{eqnarray}
one sees that \p{gnls} coincides
with eqs. \p{newgnls}, up to
change of sign of $t_2$, and admit the involution \p{nnconj}. Thus, in
addition to the transformation \p{n3}, relating eqs. \p{newgnls} to \p{gnls},
there exists the transformation \p{nn3}, which relate them up to the
$t_2$-sign. Actually, there are two more sets of such transformations
which can be derived by applying in consecutive order the involutions
\p{conj} and \p{nnconj} to the transformations \p{n3} and \p{nn3},
\begin{eqnarray}
J^{T}=\frac{1}{2}{\overline \B}(\frac{1}{2} \B {\overline \B}+
\frac{1}{2} S {\overline S} - \partial){\overline \B}~^{-1}, \quad
{\Phi}^{T} = -\frac{i}{\sqrt{2}} D ({\overline S}~{\overline \B}~^{-1}),
\quad {\overline \Phi}^{T}=\frac{i}{\sqrt{2}} \overline D ({\overline \B}S),
\label{nnn3}
\end{eqnarray}
\begin{eqnarray}
J=-\frac{1}{2}[{\overline D}\F]^{-1} (\frac{1}{2} \F {\overline \F} +
\frac{1}{2} S {\overline S} - \partial)[{\overline D}\F], \quad
\Phi=\frac{i}{\sqrt{2}}D([{\overline D}\F]^{-1}{\overline D}S), \quad
{\overline \Phi}= \frac{i}{\sqrt{2}}{\overline D}
({\cal I}^{2}{\overline S} \F),
\label{nnnn3}
\end{eqnarray}
respectively.

{}~

\noindent{\bf 5. Discrete symmetries of the $N=2$ super $(k|n,m)$-MGNLS
hierarchy.}
Following the method developed in \cite{s2,s1} and using the results of
section 4 one can easily derive the discrete symmetries of the $N=2$
$(k|n,m)$-MGNLS hierarchy. We recall that the discrete symmetries of any
integrable system represent as a rule integrable lattice dynamical systems
\cite{dls,ls,bs2}. Without going into details, let us present
only the  results. We refer to the Lax operator \p{suplax1} and quote
a basic {\it proposition}.
If the matrices $\{ B_j,{\overline B}_j,
S_j,{\overline S}_j\}$ labeled by index $j$ ($j \in {\cal Z}$) at 
some given $j$ form a solution of the $N=2$ super $(k|n,m)$-MGNLS
hierarchy, then the matrices $\{ B_{j+1},{\overline B}_{j+1},
S_{j+1},{\overline S}_{j+1}\}$ also form a solution provided they are connected
with the former ones by the following relations:
\begin{eqnarray}
&& \B_{j+1}{\overline \B}_{j+1}+
S_{j+1} {\overline S}_{j+1}-
(\B_{j+1}{\overline \B}_{j})(\B_{j} {\overline \B}_{j} +
S_{j} {\overline S}_{j})(\B_{j+1}{\overline \B}_{j})^{-1} =
2 (\B_{j+1}{\overline \B}_{j})~'~(\B_{j+1}{\overline \B}_{j})^{-1},\nonumber\\
&& {\overline D}~({\overline \B}_j S_{j}+i {\B_{j+1}}^{-1} S_{j+1})=0, \quad
D~ ({\overline S}_{j+1}\B_{j+1}+i{\overline S}_{j}{{\overline \B}_j}^{-1})=0,
\label{n6}
\end{eqnarray}
for the case of bosonic matrices $B_{j}\equiv \B_j$ and 
${\overline B}_{j}\equiv {\overline \B}_j$, or
\begin{eqnarray}
&& D ({\overline \F}_{j} S_{j}-
i[\overline D \F_{j+1}]^{-1}{\overline D} S_{j+1})=0,\quad
{\overline D}({\cal I}^{2}{\overline S}_{j+1}\F_{j+1}+
i D{\overline S}_{j} [D {\overline \F}_{j}]^{-1})=0, \nonumber\\
&& \F_{j+1}{\overline \F}_{j+1}+S_{j+1} {\overline S}_{j+1}-
([{\overline D} \F_{j+1}] [D{\overline \F}_{j}])(\F_{j} {\overline \F}_{j} +
S_{j} {\overline S}_{j})
([{\overline D} \F_{j+1}][D{\overline \F}_{j}])^{-1} \nonumber\\
&& =([{\overline D} \F_{j+1}][D{\overline \F}_{j}])~'~
([{\overline D} \F_{j+1}][D{\overline \F}_{j}])^{-1},
\label{nn6}
\end{eqnarray}
for the case when they are fermionic, $B_{j}\equiv \F_j$ and 
${\overline B}_{j}\equiv {\overline \F}_j$. The
relations \p{n6} and \p{nn6} represent the matrix generalization of a
wide class of $N=2$ supersymmetric generalized Toda--lattice equations,
constructed in \cite{s1}. The detailed analysis of them is out the scope
of the present letter and will be discussed elsewhere. Let us only mention
that among these systems one has the minimal $N=2$ supersymmetric
generalization of the bosonic non-abelian Toda lattice equations
\cite{bmrl} as well as 
one important property of theirs, i.e. the involution ${\sigma}_l$ 
($l\in {\cal Z}$) defined by
\begin{eqnarray}
&&\sigma_l B_j {\sigma}^{-1}_l=i{\overline B}_{l-j}^{T}{\cal I}, \quad
\sigma_l {\overline B}_j {\sigma}^{-1}_l= i{\cal I} B_{l-j}^{T}, \nonumber\\
&& \sigma_l S_j {\sigma}^{-1}_l=i{\overline S}_{l-j}^{T}{\cal I}, \quad
\sigma_l {\overline S}_j {\sigma}^{-1}_l= i{\cal I} S_{l-j}^{T},\nonumber\\
&&\sigma_l {\theta} {\sigma}^{-1}_l= {\overline \theta}, \quad
\sigma_l {\overline \theta} {\sigma}^{-1}_l= {\theta}, \quad
\sigma_l z {\sigma}^{-1}_l=z, \quad \sigma_l i^{*}{\sigma}^{-1}_l=-i
\label{auto1}
\end{eqnarray}
which is very useful for analyzing them.

{}~

\noindent{\bf 6. Reductions of the $N=2$ supersymmetric matrix KP
hierarchy.}
The generic form of the $N=2$ matrix KP Lax operator is\footnote{For
details concerning bosonic matrix KP and matrix (extended) Gelfand-Dickey
hierarchies, see the recent papers \cite{kl,b,fm} and references therein.}   
\begin{eqnarray}
&& L_{KP}=I{\partial} + {\sum}_{j=-\infty}^{0}(a_j + 
\omega_j D+ {\overline \omega}_j {\overline D}
+ b_j [D,\overline D])\partial^j,\label{KP}
\end{eqnarray}
where $a_j,b_j$ ($\omega_j, {\overline \omega}_j$) are generic bosonic
(fermionic) square matrix $N=2$ superfields. 
The Lax operators \p{suplax} and \p{n4}, introduced above, represent reductions
of the $N=2$ matrix KP hierarchy, characterized by a finite number of 
superfields. In the following we want to show that these examples are 
particular cases of an infinite class of reductions (with a finite number 
of superfields).

The idea of the construction of this class is based on previous work.
 
In \cite{kst,bks} a new type of $N=2$ supersymmetric
pseudo-differential Lax operators ${L}$ was introduced. It
was characterized by a non--standard $N=2$
super--residue equal to the $N=2$ superfield integral of the constant part of 
the
operator\footnote{Let us recall that standard $N=2$ super-residue is 
defined as
the $N=2$ superfield integral of the coefficient of the operator 
$[D,{\overline D}]{\partial}^{-1}$.}. In 
\cite{s1} it was observed that all these Lax operators  
possess the following important property: they commute with
one of the two fermionic covariant derivatives, $[D,L]=0$ or 
$[{\overline D},L]=0$. In other words, 
in \cite{s1} it was established that actually they are not $N=2$, but  
$N=1$ supersymmetric pseudo-differential operators. 
Consequently, the residue of these operators coincides with the
residue of $N=1$ supersymmetric pseudo-differential
operators, i.e. with the $N=1$ superfield integral of the coefficient 
of the operator 
${\overline D} {\partial}^{-1}$ or $D {\partial}^{-1}$. This would seem 
to lead to $N=1$
supersymmetric hierarchies, rather then to $N=2$ ones. Nevertheless  
the coefficients of such operators are expressed in a special way in terms
of $N=2$ superfields and their fermionic derivatives, and, as a 
result, this fact leads to $N=2$ supersymmetric systems. 
Due to commutativity of the Lax operator and the fermionic derivative, 
the $N=1$ super--residue coincides with
the $N=2$ superfield integral of the constant part of the operator 
(see eq.\p{l2}), and, 
it reproduces and justifies the
definition of the super-residue given in \cite{kst,bks} for the case of
degenerated Lax operators\footnote{In the case when the $N=1$ super--residue 
vanishes, one can use the bosonic residue to construct the Hamiltonians, 
see \cite{bks}.}. 

So, the first property of the reduced Lax operator we want to maintain
is commutativity with the fermionic derivative.
The constraint $[D, L_{KP}]=0$ for the generic $N=2$ matrix
KP Lax operator \p{KP}, can be be solved in general and takes the form
\begin{eqnarray}
&& L_{KP}^{red}=I{\partial} + a_0+ \omega_0D + 
{\sum}_{j=-\infty}^{1}(a_j{\partial} - 
[Da_j]{\overline D}+\omega_j D\partial -{1\over 2} 
 [D\omega_j][D,\overline D])\partial^{j-1}, \label{l2}
\end{eqnarray}
where $a_0$ and $\omega_0$ are chiral superfields. This Lax operator 
defines a reduction of KP with an infinite number of matrix superfields.
     
In addition in \cite{bks} the bosonic limits of the Lax operators corresponding 
to the
$N=2$ supersymmetric integrable hierarchies with $N=2$ $W_{s+1}$ second
Hamiltonian structure were conjectured. The third input are
the results we obtained in the previous sections for the 
$N=2$ supersymmetric $(k|n,m)$--MGNLS hierarchy. 

Based on these three inputs, we are lead to the following conjecture 
for the expression of
the matrix--valued pseudo--differential operator with a finite number
of superfields, representing a 
reduction of $N=2$ matrix KP hierarchy,
\begin{eqnarray}
&& (L_{KP}^{red})^s\equiv L_s=I{\partial}^{s} + 
{\sum}_{j=1}^{s-1}(J_{s-j}{\partial} - 
[DJ_{s-j}]{\overline D}){\partial}^{j-1}- J_s -
{\overline D}{\partial}^{-1}[DJ_{s}] - F {\overline F} -
F {\overline D}{\partial}^{-1} [D{\overline F}], \nonumber\\
&& [D,L_s]=0, \quad {\textstyle{\partial\over\partial t_p}}L_s =
[(L_s^{p/s})_{\geq 1},L_s],
\quad\quad H_p=\int dZ {\cal H}_p \equiv \int dZ~tr(L_s^{p/s})_{0}. 
\label{l1}
\end{eqnarray}
Here, $s \in {\cal N}$, 
the $J_j$ are $k \times k$ matrix--valued functions with the scaling
dimension in length $[J_j]=-j$, and $[F]=[{\overline F}]=-s/2$. 
One can easily verify that its bosonic limit in the scalar case, i.e., at
$k=1$, in fact reproduces the Lax operator $L^{(2)}_{[s;{\alpha}]}$ 
($L^{(3)}_{[s;{\alpha}]}$) conjectured in \cite{bks} at 
$F=\overline F=0$ ($F=\overline F=J_s=0$). 

In fact, before going into a more detailed discussion of the Lax 
operator \p{l1} in the matrix case, it is instructive to examine it in the 
simpler and more studied scalar case (i.e., $k=1$).
To start with let us mention that in
the scalar case the Lax operators \p{l1} at
$F=\overline F=J_s=0$ reproduce the scalar hierarchies first 
studied in \cite{inka} in terms of $N=1$ superfields.
One can say that in general in the scalar case the Lax operator \p{l1}
describes almost all known $N=2$ supersymmetric hierarchies. Let us present
the first few cases in explicit form.

{}~

\noindent 1. The $s=1$ scalar case.

In this case the Lax operator has the following form:
\begin{equation}
L_1=\partial- J_1-{\overline D}\partial^{-1}\left[D J_1\right] -
  F{\overline F} -F{\overline D}\partial^{-1}\left[ D {\overline F}
   \right].  \label{ll1}
\end{equation}
In the generic case this Lax operator describes the GNLS hierarchy in the KdV
basis \cite{ik,bs}. There are two possible reductions, one with 
$F={\overline F}=0$ corresponds to the $N=2$ $a=4$ KdV hierarchy \cite{kst},
the second, with $J_1=0$, corresponds to the $N=2$ GNLS hierarchy \cite{bks}.
In the special case
when we deal only with one pair of bosonic chiral-anti-chiral 
superfields $F,{\overline F}$, the Lax operator \p{ll1} describes the
$N=4$ KdV hierarchy \cite{ik}.

{}~

\noindent 2. The $s=2$ scalar case.

In this case the Lax operator
\begin{equation}
L_2=\partial^2+J_1\partial -\left[ DJ_1\right] {\overline D}
  - J_2-{\overline D}\partial^{-1}\left[D J_2\right] -
  F{\overline F} -F{\overline D}\partial^{-1}\left[ D {\overline F}
   \right]  \label{ll2}
\end{equation}
includes spin 1 superfields -- a general superfield $J_1$ and 
chiral/anti--chiral $F,{\overline F}$ ones, as well as the spin 2 general 
superfield $J_2$.
This operator has been considered in \cite{p}. It describes the extension of
the $N=2$, $a=-2$ super Boussinesq hierarchy, because in the case
of $F={\bar F}=0$ the operator \p{ll2} corresponds just to the
$a=-2$ Boussinesq hierarchy. Another possible reduction, $J_2=0$, gives
us the Lax operator for the extended quasi--$N=4$ KdV hierarchy \cite{DGI}, 
while the maximal possible reduction, $J_2=F={\overline F}=0$, reproduces the
Lax operator for the $N=2$ $a=-2$ KdV hierarchy \cite{lm}.

{}~

\noindent 3. The $s=3$ scalar case.

The last case we are going to consider explicitly, corresponds to the
$N=2$ supersymmetric extension of the bosonic system involving also a spin 4
field -- i.e. the bosonic 4--KdV hierarchy. The corresponding Lax operator
\begin{equation}
L_3=\partial^3+J_1\partial^2 -\left[ DJ_1\right] {\overline D}\partial+
J_2\partial -\left[ DJ_2\right] {\overline D}
  - J_3-{\overline D}\partial^{-1}\left[D J_3\right] -
  F{\overline F} -F{\overline D}\partial^{-1}\left[ D {\overline F}
   \right]  \label{ll3}
\end{equation}
includes general superfields $(J_1,J_2,J_3)$ with spin $(1,2,3)$,
respectively, and chiral/anti--chiral superfields $F,{\overline F}$
with spin 3/2. The generic Lax operator \p{ll3} describes
one of the three integrable systems possessing the $N=2$ $W_4$ algebra 
as the second Hamiltonian structure (in the limit $F={\overline F}=0$) \cite{yw}
and it is presented here for the first
time. The reduction $J_3=0$ gives the Lax operator for the extension of the
$N=2$ $a=-1/2$ Boussinesq hierarchy \cite{ik}, while the maximal reduction
$J_3=F={\overline F}=0$ reproduces the $N=2$ $a=-1/2$ Boussinesq hierarchy
itself \cite{bikp}.

Let us finish this excursion in the scalar $N=2$ KP reductions with
two comments which hold also in the matrix case.

First of all, it comes as a surprise that two out of three families of
integrable hierarchies with $N=2$ $W_s$ algebra as the second Hamiltonian 
structure are naturally combined in the Lax operators \p{l1} -- 
one is given by the
$L_s$ operator itself while the second is the $J_{s+1}=0$ reduction of
the $L_{s+1}$ operator (in the limit $F={\overline F}=0$). 
The remaining family of hierarchies, which starts from 
$N=2$ $a=5/2$ Boussinesq equation, has a completely different Lax operator
whose closed form is unknown yet, see for details \cite{bks}.

As a second remark, we would like to stress that the gauge transformations 
\p{n2},\p{nn2}
give the possibility to recover higher spin superfield $J_s$ in the
Lax operator $L_s$ \p{l1} starting from the reduced case $J_s=0$.
As a consequence the superalgebras which are the
second Hamiltonian structures for these hierarchies (the reduced and unreduced
ones) are closely related.
In other words the superalgebra with the currents $(J_1,\ldots ,J_s)$
and $n+m$ pairs of superfields $(F,{\overline F})$ can be 
realized in terms of the superalgebra formed by the supercurrents 
$(J_1,\ldots ,J_{s-1})$ and
$n+m+1$ pairs of $(F,{\overline F})$.
The first example of such relations between $N=2$ NLS and $N=2$
$a=4$ KdV hierarchies was elaborated in \cite{SK}.

Let us now return to the matrix hierarchies.
In what follows we verify that the operator \p{l1} is actually a 
reduction of the $N=2$ matrix KP hierarchy in the simplest cases 
corresponding to $s=1$, $s=2$ and $s=3$, by showing that
the second flows are consistently produced via the Lax pair representation
\p{l1}.

{}~

\noindent{1. The $s=1$ matrix case.}

It is obvious that in this case the operator \p{l1} reproduces 
the Lax operators \p{suplax} and \p{n4}\footnote{This correspondence
can be easily established after obvious transformation to the new basis 
in the space of the superfields.}.

{}~

\noindent{2. The $s=2, p=2$ matrix case.}
\begin{eqnarray}
&& {\textstyle{\partial\over\partial t_2}}{J_1} =
2(J_2 + F{\overline F})~'+[J_1,J_2+F{\overline F}], \nonumber\\
&& {\textstyle{\partial\over\partial t_2}}J_2 =[D,{\overline D}]J_2~'-
J_1J_2~'+ [DJ_1]{\overline D}J_2 - [DJ_2]{\overline D}J_1-
[J_1,{\overline D}DJ_2], \nonumber\\
&& {\textstyle{\partial\over\partial t_2}}F =
-F~''+ D(J_1\overline D F), \quad
{\textstyle{\partial\over\partial t_2}}{\overline F} =
{\overline F}~''+ \overline D ([D{\overline F}] J_1), \nonumber\\
&& {\cal H}_2=tr(J_2+F{\overline F}).
\label{c1}
\end{eqnarray}
The second flow equations, in the limit $F=\overline F=0$, provides the 
matrix extension of the $N=2$ $a=-2$ Boussinesq equation.

{}~

\noindent{3. The $s=3, p=2$ matrix case.}
\begin{eqnarray}
&& {\textstyle{\partial\over\partial t_2}}{J_1} =
(J_1~'+\frac{1}{3}J_1^2-2J_2)~'+\frac{1}{3}[J_1,J_1~'-2J_2], \nonumber\\
&&  {\textstyle{\partial\over\partial t_2}}J_2 = (2J_3-J_2~'
+\frac{2}{3}J_1~''+2F{\overline F})~'+\frac{2}{3}(J_2J_1~'-J_1J_2~'~
+J_1J_1~'' \nonumber\\
&& -[DJ_1]{\overline D}J_1~'+[DJ_1]{\overline D}J_2-[DJ_2]{\overline D}J_1
+[J_1,J_3+F{\overline F}]), \nonumber\\
&&  {\textstyle{\partial\over\partial t_2}}J_3 =[D,{\overline D}]J_3~'
+ \frac{2}{3}([DJ_1]{\overline D}J_3 -
[DJ_3]{\overline D}J_1-[J_1,{\overline D}DJ_3]-J_1J_3~'), \nonumber\\
&& {\textstyle{\partial\over\partial t_2}}F =
-F~''+\frac{2}{3} D(J_1\overline D F), \quad
{\textstyle{\partial\over\partial t_2}}{\overline F} =
{\overline F}~''+\frac{2}{3} \overline D ([D{\overline F}] J_1), \nonumber\\
&& {\cal H}_2=tr(J_2-\frac{1}{6}J_1^2), \quad
{\cal H}_3=tr(J_3+F{\overline F}).
\label{c2}
\end{eqnarray}
The second flow equations, in the limit $J_3=F=\overline F=0$, define
the matrix extension of the $N=2$ $a =-{1\over 2}$ Boussinesq equation.

 Let us make a few remarks about these hierarchies.
The equations \p{c1} admit the involution 
\begin{eqnarray}
&& F^{*}= i^s{\overline F}^{T}{\cal I}, \quad
{\overline F^{*}}=i^s{\cal I} F^{T}, \quad
J_j^{*}=(-1)^{j}J_j^{T}, \nonumber\\
&& {\theta}^{*}={\overline {\theta}}, \quad
{\overline {\theta}^{*}}={\theta}, \quad
t^{*}_p= (-1)^{p+1} t_p, \quad z^{*}= z, \quad i^{*}=-i,
\label{conjs}
\end{eqnarray}
for $s=2$. The same involution property is not satisfied for $s=3$.
Nevertheless 
one can still think that there exists a basis in the space of the
superfield matrices where the involution \p{conjs} is admitted for
any given value of the parameter $s$. Indeed, taking as an example 
the $s=3$ equations \p{c2}, let us introduce a new basis with the
superfield $J_2$ being replaced by
\begin{eqnarray}
J_2 \Rightarrow J_2 - \frac{1}{6}J_1^2 -\frac{1}{2}J_1~',
\label{trn}
\end{eqnarray}
while all the other superfields are unchanged.
It is a simple exercise to verify that in this new basis the involution
property \p{conjs} is satisfied.

  From eqs. \p{c1} and \p{c2} we can explicitly see that  
it is consistent to set either the superfield $J_s=0$ or 
$F=\overline F=0$ or both simultaneously. These additional reduction
properties are general, and, together with the involution properties of the
superfield matrices they can be used to straightforwardly
derive discrete symmetries of the Lax operators $L_s$ \p{l1} at
$J_s=0$ and generic $s$, following the method developed in \cite{s2,s1} 
and used in sections 4 and 5 for the case of $s=1$.

To close this Letter let us remind that there is an alternative
description of the reductions of the scalar $N=2$ supersymmetric KP
hierarchies based on Lax operators obeying the chirality
preserving condition $DL=L{\overline D}=0$, as opposed to the
condition $\left[ D,L\right]=0$ we used here. This type of Lax operators
has been introduced in \cite{p1} and then it was studied in details in 
\cite{dg}, 
where a wide class of $N=2$ scalar KP reductions was derived. One may wonder
whether these Lax operators also admit a matrix generalization.
The answer is positive and the construction is straightforward: one simply
replaces in the Lax operators of \cite{dg} superfield functions by matrix 
valued superfields: 
\begin{equation}
L_s=D\left( I\partial^{s-1}+\sum_{j=0}^{s-2} J_j\partial^j+F\partial^{-1}
     {\overline F}\right){\overline D} \label{DG1}
\end{equation}
and
\begin{equation}
{\widetilde L}_s=D\left( I\partial^{s-1} +\sum_{j=0}^{s-2}J_j\partial^j+
{\overline D}\partial^{-1}(J_s+{\overline F}\partial^{-1}F )
\partial^{-1}D\right){\overline D}, \label{DG2}
\end{equation}
where the $J_j$ are $k\times k$ matrix-valued general $N=2$ superfields,
and $F,{\overline F}$ chiral/anti--chiral rectangular matrix superfields.
The formulae for flows and Hamiltonians are the standard ones.
The two operators above are not independent but gauge related.

We have checked that all particular cases explicitly presented in \cite{dg}
admit a matrix 
generalization with Lax operators \p{DG1},\p{DG2} and
the corresponding flow equations and Hamiltonians coincide with
those which come from  the Lax operator \p{l1}. But
a rigorous proof for generic $s$ is still lacking.

{}~

\noindent{\bf Summary}.
In this paper we have defined the $N=2$ generalization of the matrix 
GNLS hierarchies, referred to as $N=2$ $(k|n,m)$--MGNLS hierarchies. 
Many results are simply stated without proof, since they are a 
straightforward generalization of the analogous results for the 
corresponding scalar hierarchies -- but this is not a general rule. We have
first introduced the Lax operator realization of the $N=2$ $(k|n,m)$--MGNLS 
hierarchies. Then we have explicitly calculated their second Hamiltonian 
structure as well as their conserved charges.
Then we have produced a new basis for the matrix superfields, the so--called
KdV basis, and defined the new hierarchies in this new basis, which also
admit local flows. Furthermore we have spelled out the discrete symmetries
of the $N=2$ $(k|n,m)$--MGNLS hierarchies, which lead to the matrix analogs
of the $N=2$ Toda lattice hierarchies. Finally we have analyzed the possibility
of viewing the previously introduced hierarchies as reductions of the
matrix $N=2$ KP hierarchy. This has led us to conjecture the existence of an
infinite family of $N=2$ matrix hierarchies, which are reductions of the
KP hierarchy and are characterized by a finite number of fields. The family
is parametrized by a natural number $s$. For $s=1$ the hierarchy corresponds to
the $N=2$ $(k|n,m)$--MGNLS one. We have verified this conjecture for the first 
few cases. We have also derived the involution and reduction properties of these
few cases and conjectured that they hold in general. 

{}~

\noindent{\bf Acknowledgments.}
We would like to acknowledge discussions we had with G. Marmo and G. Vilasi.
 S.K. and A.S. would like to thank SISSA for the hospitality and for 
financial support.
This work was partially supported by the Russian Foundation for Basic
Research, Grant No. 96-02-17634, RFBR-DFG Grant No. 96-02-00180, INTAS
Grant No. 93-1038, INTAS Grant No. 93-127 ext., INTAS Grant No. 94-2317 and 
by the EC TMR Programme, Grant FMRX-CT96-0012.

\end{document}